\documentclass[12pt]{iopart}

\usepackage{graphicx}

\eqnobysec

\begin{document}

\title[Hamiltonian walks on fractals]
{Hamiltonian walks on Sierpinski and $n$-simplex fractals}

\author{J Staji\' c\dag, S Elezovi\' c-Had\v zi\' c\ddag}

\address{\dag\
Theoretical Division and Center for Nonlinear Studies, Los Alamos
National Laboratory, Los Alamos, NM 87545, USA}
\address{\ddag\ Faculty of Physics, University of Belgrade, P.O. Box 368,
11001 Belgrade, Serbia and Montenegro}

\eads{jstajic@lanl.gov, suki@ff.bg.ac.yu}

\begin{abstract} We study Hamiltonian walks (HWs) on Sierpinski
and $n$--simplex fractals. Via numerical analysis of exact
recursion relations for the number of HWs we calculate the
connectivity constant $\omega$ and find the asymptotic behaviour
of the number of HWs. Depending on whether or not the polymer collapse
transition is possible on a studied lattice, different scaling
relations for the number of HWs are obtained. These relations are
in general different from the well-known form characteristic of
homogeneous lattices which has thus far been assumed to hold for
fractal lattices too. \end{abstract}

\submitto{\JPA}


\maketitle

\section{Introduction}

Enumeration of Hamiltonian walks (HWs), i.e. self-avoiding walks
(SAWs) that visit every site of a given lattice,
is a classic problem in graph theory, but it also has an
important role in the study of the configurational statistics of
polymers. HWs are used to model collapsed polymers
\cite{Vanderzande}, polymer melting \cite{Bascle,Jacobsen}, as
well as protein folding \cite{Camacho,Lua}. The number of all
possible HWs on a lattice is related to the configurational
entropy of a collapsed polymer system, and also to the optimal
solutions to the traveling salesman problem \cite{salesman,Dean}.
Enumeration of HWs, closed or open, is a difficult combinatorial problem, which has
been exactly solved only for few lattices, namely, the
two-dimensional Manhattan oriented square lattice
\cite{Kasteleyn,Malakis}, the two-dimensional ice lattice
\cite{Lieb}, the two-dimensional hexagonal lattice
\cite{Suzuki,Batchelor1}, the Sierpinski gasket fractal
\cite{salesman}, and the 4-simplex fractal \cite{Bradley}. The
number of HWs has also been calculated numerically for various
lattices by means of direct enumeration \cite{Mayer,Pande},
transfer matrix methods \cite{Schmalz,Kloczkowski,Foster}, and
Monte Carlo estimates \cite{Jaeckel,Hsu}. The field theory
representation for this problem was introduced in \cite{Orland},
and further developed in \cite{Jane0} and \cite{Higuchi1}.

{The purpose of this paper is to understand the topological
properties of Hamiltonian walks on several families of fractal
lattices.} In order to do this we study the asymptotic behavior of
the number of closed HWs $C_N$ for a large number of vertices $N$.
This analysis yields the values of the so-called connectivity
constant $\omega$ which has the physical meaning of the {average
number of steps available to the walker having already completed
 a large number of
steps}. It also provides insight into the spatial
distribution of walks present at large $N$; this can be related to
detailed studies of knot delocalisation in Refs. \cite{Orlandini}.

The number $C_N$ is for homogeneous lattices with $N \gg 1$ expected to
take the form
\begin{equation}
C_N\sim \omega^N {\mu_S}^{N^\sigma}N^a\, . \label{eq:asimptotika}
\end{equation}
Here $\omega$ is the connectivity constant and the term with $\mu_S$
represents a surface correction
($\mu_S<1$), with $\sigma=(d-1)/d$ ($d$ being the dimensionality
of the lattice). This differs from the ordinary SAW case, where no
surface term ${\mu_S}^{N^\sigma}$ is expected, i.e. the number of
SAWs of length $N$ behaves as $\mu^N N^a$  for large $N$.
Furthermore, the exponent $a$ is universal in the SAW case, i.e. it
depends only on the dimensionality of the lattice, whereas for HWs
it may depend on other, not yet identified characteristics of the
lattice. To the lowest approximation in Eq.~(\ref{eq:asimptotika})
$C_N \sim \omega^N$, and the connectivity constant can be defined
as
\begin{equation}
 \ln\omega  =\lim_{N\to\infty}{{\ln C_{N}}\over  N}\, .
 \label{eq:definicija}
\end{equation}

  For a better understanding of these problems it is
helpful to study HWs on fractal lattices. As was first recognized by
Bradley \cite{Bradley}, the self-similarity of fractal lattices is a
useful tool for the exact and computationally fast iterative
generation and enumeration of all HWs on an unlimitedly large
corresponding fractal structure. In this paper we extend Bradley's
algorithm to two-- and three--dimensional Sierpinski fractal
families, as well as $n$--simplex fractals with $n>4$.
We consider several families of fractals in order to compare the
obtained results and be able to draw more general conclusions
about the character of the Hamiltonian walks on different classes
of lattices.
In the case
of the two--dimensional Sierpinski fractal family an exact closed
form result for the connectivity constant is obtained {due to the
simple form of the recursion relations for the numbers of HWs}. For
the three-dimensional Sierpinski fractals which can model physically
more frequently encountered systems a numerical approach is
necessary. The study of asymptotics of the number of HWs shows that
the surface term in Eq.~(\ref{eq:asimptotika}) only appears for
fractals on which the collapse transition from the polymer coil to
the globule phase is possible. This is the same class of lattices
for which delocalised HWs dominate over localised ones for large $N$.

 The paper is organized as follows. In section 2 we describe the two--dimensional
Sierpinski fractal family and obtain exact recursion relations for
the number of HWs, as well as the closed exact formula for
$\omega$. Recursion relations for HWs on three--dimensional
Sierpinski fractals are given and analyzed in section 3. A similar
method for analyzing HWs on 5- and 6-simplex lattices is presented
in section 4. Finally, in section 5 we discuss all our findings
and related results obtained by other authors.

\section{Hamiltonian walks on two--dimensional Sierpinski fractals}

\begin{figure}
\begin{center}
\includegraphics[width=80mm]{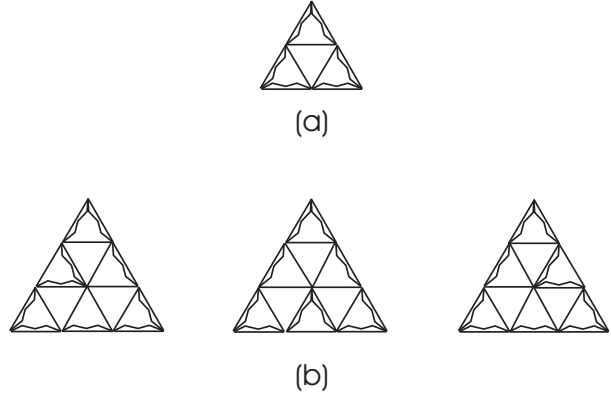}
\end{center}
\caption{The $(l+1)$th order generators $G_{l+1}(2)$ (a) and
$G_{l+1}(3)$ (b) for $b=2$ and $b=3$ two-dimensional Sierpinski
fractals, with the possible closed Hamiltonian walks
configurations depicted. The small up-oriented triangles are the
respective $l$th order generators, and the lines that traverse
them represent the open Hamiltonian walks.} \label{fig:CL}
\end{figure}

\begin{figure}
\begin{center}
\includegraphics[width=100mm]{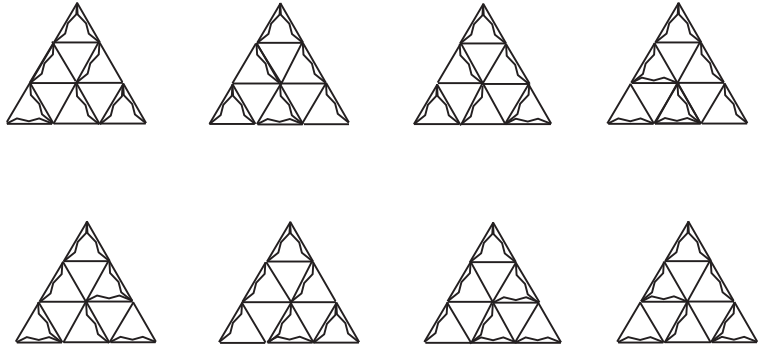}
\end{center}
\caption{ Schematic representation of the recursion relation for
the number of open Hamiltonian walks (2.3) of $h$--type on $b=3$
two-dimensional Sierpinski fractal structures.}
\label{fig:recurSG}
\end{figure}

We begin by defining the two-dimensional (2d) Sierpinski fractal family.
Each member of the 2d SF family (labeled by $b$)
can be constructed recursively, starting with an equilateral
triangle  that contains $b^2$ smaller equilateral triangles
(generator $G_1^2(b)$). The subsequent fractal stages are
constructed self--similarly, by replacing each of the $b(b+1)/2$
upward--oriented small triangles of the initial generator by a new
generator. To obtain the $l$th--stage fractal lattice $G_l^2(b)$,
which we shall call the $l$th order generator, this process of
construction has to be repeated $l-1$ times, and the complete
fractal is obtained in the limit $l\to\infty$. It is easy to see
that, for any 2d SF, each closed HW on the $(l+1)$th generator is
comprised of HWs which enter and exit $l$th order generators. Let
$C_l$ be the number of closed HWs on the $l$th order generator,
whereas $h_l$ and $g_l$  are the numbers of HWs which enter the
$l$th order generator at one vertex, and leave it at the other, with
or without visiting its third vertex, respectively. Then, it can be
shown  that a simple relation
\begin{equation}
C_{l+1}=Bh_l^\alpha g_l^\beta   \label{eq:zatvorene2dSF}
\end{equation}
is valid for $l\geq 1$ (see Appendix A). Here $B$ is a constant
that depends only on SF parameter $b$, whereas exponents $\alpha$
and $\beta$ are equal to
\begin{equation}
\alpha=b+1, \quad  \beta={{(b+1)(b-2)}\over 2 } \,
.\label{eq:eksponenti1}
\end{equation}
For instance, the explicit form of the relation
(\ref{eq:zatvorene2dSF}) for $b=2$ is $C_{l+1}=h_l^3$ and for
$b=3$: $C_{l+1}=3h_l^4g_l^2$, which is illustrated in Figure~
\ref{fig:CL}.
One can also show (Appendix A) that numbers $h_l$ and $g_l$ of
open HWs obey the following closed set of recursion relations
\begin{equation}
h_{l+1}=Ah_l^xg_l^y, \quad g_{l+1}=Ah_l^{x-1}g_l^{y+1} ,
\label{eq:otvorene2dSF}
\end{equation}
for all $l\geq 1$, where $A$ is again a constant, different for every $b$, and
\begin{equation}
x=b, \quad y={{b(b-1)}\over 2}\, . \label{eq:eksponenti2}
\end{equation}
For $b=2$ these relations have the form: $h_{l+1}=2h_l^2g_l$,
$g_{l+1}=2h_lg_l^2$ and for $b=3$:   $h_{l+1}=8h_l^3g_l^3$,
$g_{l+1}=8h_l^2g_l^4$ (see Figure~\ref{fig:recurSG}).
From the relations (\ref{eq:otvorene2dSF}) it follows straightforwardly
that
\begin{equation}
{{g_l}\over{h_l}}={{g_1}\over{h_1}}=K \, ,\label{eq:odnos}
\end{equation}
for any $l\geq 1$, so that from (\ref{eq:zatvorene2dSF}) and
(\ref{eq:eksponenti1}) one gets $C_{l+1}=BK^\beta h_l^{b(b+1)/2}$.
Since the number $N_l$ of sites on the $l$th order generator
satisfies the recursion relation
\begin{equation}
N_{l+1}={{b(b+1)}\over 2}N_l-(b^2-1) \label{eq:cvorovi},
\end{equation}
according to (\ref{eq:definicija}) it then follows that
\[
\ln\omega =\lim_{l\to\infty}{{\ln C_{l+1}}\over{N_{l+1}}}=
\lim_{l\to\infty}{{\ln h_l}\over{N_{l}}} \, .
\]
On the other hand, from (\ref{eq:otvorene2dSF}),
(\ref{eq:eksponenti2}), and (\ref{eq:odnos}) one has
$h_{l+1}=AK^yh_l^{b(b+1)/2}$, i.e. $\ln h_l$ satisfies the
difference equation $\ln h_{l+1}=\ln AK^y+{{b(b+1)}\over 2}\ln h_l$,
whose solution is
\begin{equation}
\ln h_{l+1}={{1-[b(b+1)/2]^l}\over{1-[b(b+1)/2]}}\ln(AK^y)+
\left({{b(b+1)}\over 2}\right)^l\ln h_1 \, . \label{eq:hkonacno}
\end{equation}
From this equation, together with the explicit expression for the
number of sites
\begin{equation}
N_{l+1}={{b+4}\over{b+2}}\left({{b(b+1)}\over
2}\right)^{l+1}+2{{b+1}\over{b+2}}\, , \label{eq:cvorovi2}
\end{equation}
which  follows directly from the relation (\ref{eq:cvorovi}), one can
derive the general form
\begin{equation}
\omega
=A^{\displaystyle{4\over{b(b+4)(b^2-1)}}}h_1^{\displaystyle{4\over{b(b+1)(b+4)}}}
g_1^{\displaystyle{2\over{(b+1)(b+4)}}}  \label{eq:closedform}
\end{equation}
of the connectivity constant (\ref{eq:definicija}) for HWs on
two--dimensional SFs. Consequently, in order to calculate $\omega$
for any particular 2d SF, one should find the numbers $h_1$ and
$g_1$ of open HWs on the generator, and the number $A$ of all
Hamiltonian configurations which are relevant for recursion
relations (\ref{eq:otvorene2dSF}). In Table 1
\begin{table}
\caption{Values of the number $A$ of Hamiltonian configurations
relevant for the recursion relations (2.3), numbers $h_1$ and $g_1$
of open HWs on the generators of two--dimensional Sierpinski
fractals, connectivity constant $\omega$ for HWs, and connectivity
constant $\mu$ for SAWs (obtained via RG method -
$\mu^{\mathrm{RG}}$ and numerically estimated $\mu^{\mathrm{num}}$
in [28]), for $2\leq b\leq 8$.}
\begin{indented}
\item[]
\begin{tabular}{crrrll}
\br
$b$&$A$&$h_1$&$g_1$&$\omega$&$\mu^{\mathrm{RG}}(\mu^{\mathrm{num}})$\\
\mr
 2&2 &2&3&1.31798&2.288($2.282\pm 0.007$)\\ 3&8
&10&11&1.39157& 2.491($2.49\pm 0.02$)\\ 4&40 &92&112&1.46186&
2.656($2.686\pm 0.004$)\\ 5&360  &1 852&2 286&1.52155&
2.791($2.82\pm 0.01$)\\ 6&3 872  &78 032&94 696&1.56895&2.904
($2.92\pm 0.02$)\\ 7&62 848 &6 846 876&8 320 626&1.61011&
3.005($2.99\pm 0.05$)\\ 8&1 287 840&1 255 156 712&1 527 633
172&1.64528& -------($3.13\pm 0.07$)\\ \br
\end{tabular}
\end{indented}
\end{table}
we present these numbers, together with the values of $\omega$,
for $2\leq b\leq 8$. As one can see, for $b=2$, and $b=3$ the numbers
$g_1$, $h_1$, and $A$ are small and can be directly enumerated,
whereas for larger values of $b$ they quickly increase, so that
enumeration should be computerized (calculation of the numbers
$A$, $g_1$, and $h_1$ required 13 minutes for $b=7$ case, and
about 100 hours for $b=8$, both on a computer with a processor
MIPS R10000, Rev 2.6 on 180 MHz). One should mention that
the connectivity constant for the Sierpinski gasket  ($b=2$) has
already been calculated in a different way by Bradley
\cite{salesman}.

Combining equations (\ref{eq:hkonacno}), (\ref{eq:cvorovi2}) and
(\ref{eq:closedform}) it is not difficult to see that $h_l=G
\omega^{N_l}$, where $G$ depends only on $b$, and correspondingly
$C_l\sim \omega^{N_l}$. Comparing with (\ref{eq:asimptotika}) one
can conclude that neither surface nor power correction terms are
present in the scaling form for the number of closed HWs on
two-dimensional Sierpinski fractals.

\subsection*{Comparison with the self-avoiding walk case}

It is interesting to compare the value of the connectivity constant
obtained for the case
of Hamiltonian walks to that corresponding to all possible
self-avoiding configurations on 2d SFs.
Our algorithm for enumerating HWs is easily adjusted for
that purpose.
By
means of exact renormalization group (RG) approach
\cite{Dhar,Rammal}, these configurations can be used for calculating
the connectivity constant $\mu$ for ordinary SAWs, which was done
earlier only for the $b=2$ case \cite{Rammal}. Here we extend an
exact RG calculation of the connectivity constant $\mu$ for any $b$.

The connectivity constant $\mu$ for the SAW model is equal to
$\mu=\lim_{N\to\infty}(c_{N+1}/c_N)=\lim_{N\to\infty}(p_{N+1}/{p_N})$,
where $c_N$ ($p_N$) is the average number of distinct open (closed)
$n$--step SAWs.  In order to calculate $\mu$ within the exact RG
approach, one should introduce two generating functions $B^{(l)}$
and $B_1^{(l)}$:
\[
B^{(l)}=\sum_N {\mathcal B}_N^{(l)}x^N\, ,\quad  B_1^{(l)}=\sum_N
{\mathcal B}_{1,N}^{(l)}x^N\, ,
\]
where $x$ is the statistical weight of each step of the SAW
(fugacity), whereas ${\mathcal B}_N^{(l)}$ (${\mathcal
B}_{1,N}^{(l)}$) is the number of SAWs which enter the $l$th order
generator $G_{l}^2(b)$ at one vertex, and leave it at the second,
without (with) visiting the third one. For every 2d SF lattice
functions $B$ and $B_1$ obey recursion relations of the following
form
\begin{equation}\fl
B^{(l+1)}=
\sum_{i,j}f_{i,j}(b)\left(B^{(l)}\right)^i\left(B_1^{(l)}\right)^j\,
,\qquad
B_1^{(l+1)}=\sum_{i,j}g_{i,j}(b)\left(B^{(l)}\right)^i\left(B_1^{(l)}\right)^j\,
, \label{eq:recurB}
\end{equation}
where $f_{i,j}(b)$ and $g_{i,j}(b)$ are coefficients that do not
depend on $l$, but do depend on the fractal parameter $b$. The
initial conditions are $B^{(0)}=x$, $B_1^{(0)}=x^2$ and the
connectivity constant $\mu$ is equal to $1/x^*$, where $x^*$ is the
value of the fugacity for which one approaches the fixed point
$(B^*,B_1^*)$ of (\ref{eq:recurB}), after a large (infinite) number
of iterations.

The explicit RG recursion relations for 2d SFs with $b=2$ are
\[
 \fl B'={B^2} + {B^3} + 2BB_1 + \underline{2{B^2}B_1} + {B_1^2} ,
\quad B_1' = {B^2}B_1 + \underline{2B{B_1^2}}
\]
and for $b=3$:
\begin{eqnarray}
 \fl B'=
  {B^3} + 3{B^4} + {B^5} + 2{B^6} + 3{B^2}B_1 + 12{B^3}B_1 +
   4{B^4}B_1 + 8{B^5}B_1 +3B{B_1^2} + 16{B^2}{B_1^2} \nonumber \\
\lo{+} 5{B^3}{B_1^2} + \underline{8{B^4}{B_1^2}} +
    + {B_1^3} + 8B{B_1^3} +
   2{B^2}{B_1^3} + {B_1^4} \nonumber\\
\fl B_1' = {B^4}B_1 + 2{B^5}B_1 + 4{B^3}{B_1^2} + 8{B^4}{B_1^2} +
   5{B^2}{B_1^3}+ \underline{8{B^3}{B_1^3}} + 2B{B_1^4}\nonumber
\end{eqnarray}
whereas relations for $b=4$ and 5 are given in Appendix B. For $b=6$
and 7 RG relations are too cumbersome to be quoted here, but they
are available upon request. Underlined terms in quoted RG relations
correspond to the Hamiltonian configurations, as one can check by
direct comparison with relations (\ref{eq:otvorene2dSF});
replacing $g_l$ and $h_l$ in (\ref{eq:otvorene2dSF})  by $B$ and
$B_1$, respectively, one obtains underlined terms in the RG
relations (values of coefficient $A$ are given in table 1).
 It is obvious that for larger $b$ values the number of SAW
configurations is much larger than the number of Hamiltonian
configurations, and for that reason we were not able to enumerate all SAW
configurations on 2d SF beyond $b=7$.

For all $b$ considered here recursive relations (\ref{eq:recurB})
have only one nontrivial fixed point $(B^*_b,0)$, where $B^*_b$ lies
in the interval $0<B^*_b<1$. One can check  that {by setting $B_1=0$
in (\ref{eq:recurB}) RG equations used in \cite{EKM} for
calculating critical exponent $\nu$ (connected with the mean
end-to-end distance) for SAWs on 2d SFs  are recovered}. Of course,
SAW model treated in \cite{EKM} is slightly different (each unit
triangle within the fractal can be traversed only along one side)
from the usual one, treated here, but both of them belong to the
same universality class, i.e. the critical exponent $\nu$ is equal
for both considered SAW models. This is not the case with the
connectivity constant, which is a nonuniversal quantity, so
 an extra RG parameter $B_1$ had to be introduced.

The final RG results $\mu^{\mathrm RG}$ for the SAW connectivity
constant are given in the last column of Table 1, together with the
corresponding values $\mu^{\mathrm num}$ obtained in \cite{Chalub}
using a graph counting technique. As one can expect, the connectivity
constant $\mu$ for SAW is larger  than $\omega$ for HW model, for
every considered SF, since the physical meaning of the connectivity
constant is the average number of steps available to the walker
after $N$ steps completed, for large $N$. One can also see that
the values of $\mu$ obtained by two different methods for $b=4$ and 5
are not in good agreement. The RG method applied here is exact,
implying that numerical estimations within the graph counting
technique used in \cite{Chalub} were not accurate enough.

\section{Hamiltonian walks on three--dimensional Sierpinski
fractals}\label{sec:3dSG}

We proceed by analyzing the properties of HWs on
the three--dimensional (3d) SF family.
A 3d SF can be constructed recursively as
in the 2d case, the only difference being that the generator
$G_1^3(b)$ of the fractal with the parameter $b$ is no longer a
triangle, but a tetrahedron that contains $b(b+1)(b+2)/6$ upward
oriented smaller tetrahedrons. Consequently,  one should observe
four types of open HWs that traverse the $l$--th order generator
in order to obtain the overall number of closed HWs on 3d SF. The
first three types are HWs which enter the generator at one vertex
and leave it at another, meanwhile
\begin{itemize}
\item  visiting the third, but not the fourth vertex ($g$--type),
\item  visiting both the third and the fourth vertex ($h$--type),
\item  visiting neither the third nor the fourth vertex ($i$--type).
\end{itemize}
HWs of the fourth possible type ($j$--type) consist of two
self-avoiding branches, and correspond to the walks that enter the
generator and leave it without visiting the remaining two
vertices, then, after visiting other parts of the lattice, enter
the same generator again at the third corner vertex and finally
leave it at the fourth corner vertex. Examples of these types of
walks are sketched in figure 3.
\begin{figure}
\begin{center}
\includegraphics[width=100mm]{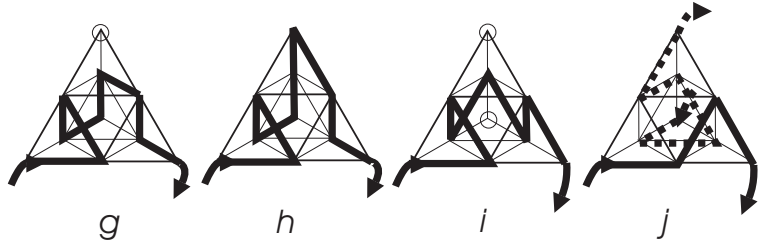}
\end{center}
\caption{Examples of four possible types of open HWs on $G_1^3(2)$. Vertices not visited by the Hamiltonian walker
are encircled.} \label{fig:3dSG}
\end{figure}
In principle, it is possible to establish a closed set of recursion relations for the numbers $g_l, h_l, i_l$, and
$j_l$ of the corresponding walks for any 3d SF in the following form
\begin{eqnarray}
 g_{l+1}&=&\sum_{n,m,k}{\mathcal G}(n,m,k)g_l^n h_l^m i_l^k
j_l^{{{b(b+1)(b+2)}\over 6}-(n+m+k)}\, ,\nonumber\\
h_{l+1}&=&\sum_{n,m,k}{\mathcal H}(n,m,k)g_l^n h_l^m i_l^k j_l^{{{b(b+1)(b+2)}\over 6}-(n+m+k)}\, ,\nonumber \\
 i_{l+1}&=&\sum_{n,m,k}{\mathcal I}(n,m,k)g_l^n h_l^m i_l^k
j_l^{{{b(b+1)(b+2)}\over 6}-(n+m+k)}\, ,\nonumber\\
j_{l+1}&=&\sum_{n,m,k}{\mathcal J}(n,m,k)g_l^n h_l^m i_l^k
j_l^{{{b(b+1)(b+2)}\over 6}-(n+m+k)}\, , \label{eq:recurSF3d}
\end{eqnarray}
where ${\mathcal G}(n,m,k)$, ${\mathcal H}(n,m,k)$, ${\mathcal
I}(n,m,k)$, and ${\mathcal J}(n,m,k)$ are the numbers of open
Hamiltonian configurations of the corresponding types, with $n$
branches of the $g$--type, $m$ branches of the $h$--type, $k$ branches of
the $i$--type, and $[b(b+1)(b+2)/6-(n+m+k)]$ branches of the $j$--type.
The number $C_{l+1}$ of all closed HWs within the $(l+1)$th order
generator for any $b$ is equal to
\[
C_{l+1}=\sum_{n,m,k}{\mathcal B}(n,m,k)g_l^n h_l^m i_l^k
j_l^{{{b(b+1)(b+2)}\over 6}-(n+m+k)}\, ,
\]
where ${\mathcal B}(n,m,k)$ is the number of all closed
Hamiltonian configurations with $n$ branches of the $g$--type, $m$
branches of the $h$--type, $k$ branches of the $i$--type, and
$[b(b+1)(b+2)/6-(n+m+k)]$ branches of the $j$--type. As one can see,
the recursion relations (\ref{eq:recurSF3d}) for the number of open
HWs on 3d Sierpinski fractals are much more complicated than the
corresponding equations (\ref{eq:otvorene2dSF}) for 2d SFs.
Consequently, it is not possible to find explicit expression,
similar to (\ref{eq:closedform}), for the connectivity constant
$\omega$. Instead, one should perform a numerical analysis of the
recursion relations (\ref{eq:recurSF3d}) in order to find the
value of $\omega$. We shall demonstrate the method on the
particular case of the $b=2$ fractal.

By computer enumeration of the possible HW configurations within
the $(l+1)$th-order generator $G_{l+1}^3(2)$ of the $b=2$ 3d
Sierpinski fractal, we found the following recursion relations:
\begin{eqnarray}
g'&=&6g^2j^2+4g^3j+2g^4+12ij^2h+24igjh+24ig^2h+8i^2h^2\, , \label{eq:g3d}
\\
 h'&=&24j^2hg+16h^2ij+16hg^3+32h^2gi+24g^2hj\, , \label{eq:h3d} \\
 i'&=&12igj^2+12ig^2j+8ig^3+8i^2jh+16i^2gh\, , \label{eq:i3d} \\
 j'&=&8i^2h^2+48igjh+22j^4+2g^4+8g^3j+24ig^2h\, ,
 \label{eq:b23dSFrecur}
\end{eqnarray}
where we have used the prime symbol as a superscript for the
numbers of HWs on the $(l+1)$th-order generator $G_{l+1}^3(2)$ and
no indices for the $l$th-order numbers. From the definition
(\ref{eq:definicija}) of the connectivity constant and the formula
for the number $C_{l+1}$ of closed HWs within $G_{l+1}^3(2)$:
\begin{equation}
C_{l+1}=16g_l^2h_l^2\, , \label{eq:chw}
\end{equation}
it then follows that
\begin{equation}
\ln\omega=\lim_{l\to\infty}{{\ln C_{l+1}}\over{N_{l+1}}}={1\over
2}\lim_{l\to\infty}{{\ln g_l}\over{N_{l}}}+{1\over
2}\lim_{l\to\infty}{{\ln h_l}\over{N_{l}}} \, , \nonumber
\label{eq:jedan}
\end{equation}
where $N_l=2(4^l+1)$ is the number of sites in $G_l^3(2)$. On the
other hand, from  the recursion relation (\ref{eq:g3d}) for the
numbers $g_l$ one obtains
\begin{eqnarray} \fl \lim_{l\to\infty}{{\ln
g_{l+1}}\over{N_{l+1}}}={1\over 2}\lim_{l\to\infty}{{\ln
g_l}\over{N_{l}}}+{1\over 2}\lim_{l\to\infty}{{\ln
j_l}\over{N_{l}}}\nonumber\\ \lo{+}{1\over 2}
\lim_{l\to\infty}{1\over{4^{l+1}}} \ln\left(6 +4\,x_l+2\,{x_l^2} +
24\,y_l\,z_l+{\frac{12\,y_l\,z_l}{{x_l^2}}}
+{\frac{24\,y_l\,z_l}{x_l}}+{\frac{8\,{y_l^2}\,{z_l^2}}{{x_l^2}}}\right)\label{pomocg}
\end{eqnarray}
where $ x_l=g_l/j_l$,  $y_l=h_l/j_l$ and $ z_l=i_l/j_l$. In a
similar way, from (\ref{eq:h3d}) it follows that:
\begin{eqnarray} \fl \lim_{l\to\infty}{{\ln
h_{l+1}}\over{N_{l+1}}}={1\over 4}\lim_{l\to\infty}{{\ln
h_l}\over{N_{l}}}+{1\over 4}\lim_{l\to\infty}{{\ln
g_l}\over{N_{l}}}+{1\over 2}\lim_{l\to\infty}{{\ln
j_l}\over{N_{l}}}\nonumber\\ +{1\over 2}
\lim_{l\to\infty}{1\over{4^{l+1}}} \ln 8\left( 3 + 3\,x_l +
2\,{x_l^2} + 4\,y_l\,z_l + {\frac{2\,y_l\,z_l}{x_l}} \right)\,
.\label{pomoch}
\end{eqnarray}
The new variables $x_l$, $y_l$ and $z_l$ fulfill recursion relations
which are easy to deduce from their definitions and Eqs.
(\ref{eq:g3d}-\ref{eq:b23dSFrecur}), and are not difficult to iterate
(starting with initial values $x_1={{11}\over{14}}$, $y_1=1$ and
$z_1={4\over 7}$, following from the numbers $g_1=88$, $h_1=112$,
$i_1=64$ and $j_1=112$, found by direct computer enumeration of the
corresponding HWs within the generator $G_1^3(2)$). One quickly
finds that $x_l$, $y_l$, and $z_l$ tend to zero, and $z_l\ll x_l\ll
y_l$, $x_l^2\sim y_l z_l$ for large $l$, meaning that the last terms on
the right--hand sides of equations (\ref{pomocg}) and (\ref{pomoch})
tend to zero. It is then straightforward to see that from
(\ref{pomocg}) and (\ref{pomoch}) {it follows that}
\[
\lim_{l\to\infty}{{\ln g_l}\over{N_l}}=\lim_{l\to\infty}{{\ln
h_l}\over{N_l}}=\lim_{l\to\infty}{{\ln j_l}\over{N_l}}\, ,
\]
and, according to (\ref{eq:jedan}), one finds
\begin{equation}
\ln\omega=\lim_{l\to\infty}{{\ln j_l}\over{N_{l}}}\, .
\label{eq:omegaprekoj}
\end{equation}
Instead of the number $j_l$, which rapidly grows with $l$, it is
convenient to introduce yet another variable
\begin{equation}
u_l={{\ln{j_l}}\over{4^l}}-
 {{\ln 22}\over 3}\left({1\over 4}-{1\over{4^l}}\right)\,
 ,\label{eq:defu}
\end{equation}
which
has the initial value $u_1={{\ln 112}\over 4}$. Numerically
iterating its recursion relation, together with those for
$x_l$, $y_l$ and $z_l$, one can show that $u_l$
tends to 1.2507788499... when $l\to\infty$. Finally, since
$\ln\omega={1\over 2}\lim_{l\to\infty}u_l+{1\over{24}}\ln 22$,
the connectivity constant for the $b=2$ 3d Sierpinski fractal is equal
to $ \omega=2.12587...$.

In order to find the first correction to the leading-order
behavior of $C_l$ one needs to know the asymptotic behavior of the
numbers $x_l$, $y_l$ and $j_l$, according to the formula
\begin{equation}
{{\ln C_{l+1}}\over{N_{l+1}}}={{\ln 16}\over{N_{l+1}}}+2{{\ln
x_l}\over{N_{l+1}}}+2{{\ln y_l}\over{N_{l+1}}}+4{{\ln
j_l}\over{N_{l+1}}}\, ,  \label{eq:prvapopravka}
\end{equation}
following from (\ref{eq:chw}) and the definition of $x_l$ and $y_l$.
Keeping only the leading-order terms in the recursion relation for
$x_l$ one gets
$x_{l+1}\approx \mathrm{const} x_l^2$, which means that $x_l$
behaves as
\begin{equation}
x_l\sim \lambda^{2^l}\label{eq:asimptotikax}
\end{equation}
for large $l$. Numerically iterating the recursion relations for
$x_l$, $y_l$ and $z_l$ one finds $\lambda=\lim_{l\to\infty}{{\ln
x_l}\over{2^l}}=0.9055...$. It then follows that $y_{l+1}\approx
{{12}\over{11}}x_ly_l$, $z_{l+1}\approx {6\over{11}}x_lz_l$,
implying that ${{y_l}\over{z_l}}\sim 2^l$, which, together with the
numerically established relation $x_l^2\sim y_l z_l$, gives
\begin{equation}
z_l\sim 2^{-l/2}x_l\sim 2^{-l/2}\lambda^{2^l}\, , \quad y_l\sim
2^{l/2}\lambda^{2^l}\, .\label{eq:asimptotikayz}
\end{equation}
On the other hand, from the definition of $u_l$ (\ref{eq:defu}) and
the corresponding recursion relation it follows
\begin{eqnarray}
\fl {{\ln{j_l}}\over{4^l}}=u_1+\sum_{k=1}^{l-1}{1\over{4^{k+1}}}\ln\left(1 + {4\over{11}}{{x_k}^3} +
{1\over{11}}{{x_k}^4} + {{24}\over{11}}{x_k}{y_k}{z_k} + {{12}\over{11}}{{x_k}^2}{y_k}{z_k} +
     {4\over{11}}{y_k}^2{z_k}^2\right)\nonumber\\
     \lo{+} {{\ln 22}\over 3}\left({1\over 4}-{1\over{4^l}}\right)\nonumber\, .
\end{eqnarray}
Then, using (\ref{eq:omegaprekoj}), one can write
\begin{eqnarray} \fl
{{\ln{j_l}}\over{4^l}}=2\ln\omega-{1\over{4^l}}{{\ln{22}}\over 3}\nonumber\\
\lo{-}\sum_{k=l}^{\infty}{1\over{4^{k+1}}}\ln\left(1 + {4\over{11}}{{x_k}^3} + {1\over{11}}{{x_k}^4} +
{{24}\over{11}}{x_k}{y_k}{z_k} + {{12}\over{11}}{{x_k}^2}{y_k}{z_k} +
     {4\over{11}}{y_k}^2{z_k}^2\right)\, , \nonumber
\end{eqnarray}
and consequently, since
\begin{eqnarray}\fl
\sum_{k=l}^{\infty}{1\over{4^{k+1}}}\ln\left(1 + {4\over{11}}{{x_k}^3} + {1\over{11}}{{x_k}^4} +
{{24}\over{11}}{x_k}{y_k}{z_k} + {{12}\over{11}}{{x_k}^2}{y_k}{z_k} +
     {4\over{11}}{y_k}^2{z_k}^2\right)\nonumber\\
     \lo{\leq}\ln{{104}\over{11}}\sum_{k=l}^\infty{1\over{4^{k+1}}}={1\over
     3}{1\over{4^l}}\ln{{104}\over{11}}\, , \nonumber
\end{eqnarray}
which is not difficult to show, one obtains
\begin{equation}
\ln j_l=2*4^l\ln\omega+O(1)\, . \label{eq:asimptotikaj}
\end{equation}
Finally, from (\ref{eq:prvapopravka})--(\ref{eq:asimptotikaj})
it follows that
\[
\ln C_l=N_l\ln\omega+N_l^{1/2}\ln\lambda^{\sqrt 2}+{1\over 2}\ln N_l+O(1)\, ,
\]
which means that the behavior (\ref{eq:asimptotika}) of the number of
HWs, expected for homogeneous lattices, is also satisfied for this
fractal lattice, with the following values of the exponents:
\[
\sigma={1\over 2}\, , \quad a={1\over 2}\, .
\]

The number of all possible HW configurations within a generator of
the 3d Sieprinski fractal  grows rapidly with $b$. In
\ref{Dodatak3dSF} we give the corresponding recursion relations for
the numbers $g_l$, $h_l$, $i_l$ and $j_l$ found for the $b=3$ 3d SF,
together with their initial values, and the formula for the number
$C_{l+1}$ of closed HWs. The CPU time required for the enumeration and
classification of HW configurations was  so
long for the $b=3$ case that we could not go beyond it. However, the method used for
$b=2$ 3d SF in principle could be applied for any $b$ with no
qualitative difference. Analyzing recursion relations given in
\ref{Dodatak3dSF} in that manner,  for the $b=3$ 3d SF we found that
the number $C_l$ of closed HWs obeys a scaling form similar to
that of the $b=2$ 3d case with the following values for the connectivity
constant $\omega$ and exponents $\sigma$ and $a$:
\[ \omega=2.2722364...\, , \quad
\sigma={{\ln 3}\over{\ln 10}} \, , \quad a =0.386... \, .\]

\section{Hamiltonian walks on $\mathbf n$--simplex fractals}

To complete our analysis of the nature of Hamiltonian walks on
2d and 3d Sierpinski fractal lattices, we now turn to n-simplexes,
which are in some ways a generalization of SFs for $b=2$ in $n-1$ dimensions.
To obtain an $n$-simplex lattice \cite{Dhar} one starts with a
complete graph of $n$ points and replaces each of these points by
a new complete graph of $n$ points. The subsequent stages are
constructed self-similarly, by repeating this procedure. After $l$
such iterations one obtains an {\it $n$-simplex of order $l$},
whereas the complete $n$-simplex lattice is obtained in the limit
$l\to\infty$.
\begin{figure}
\begin{center}
\includegraphics[width=80mm]{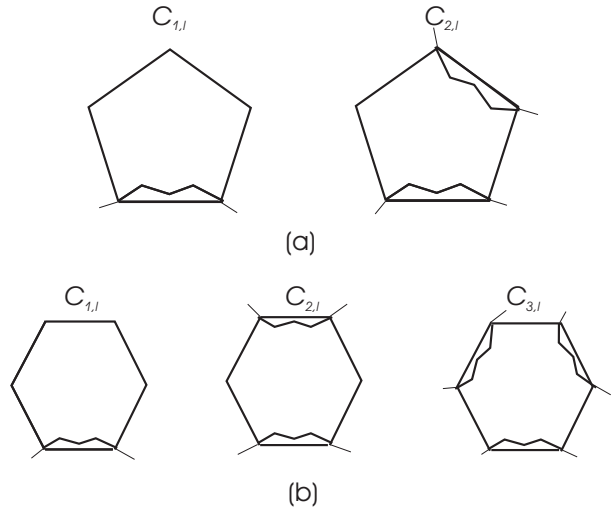}
\end{center}
\caption{ Schematic representation of types of open Hamiltonian walks through (a) 5-simplex and (b) 6-simplex of
order $l$.} \label{fig:5simplex}
\end{figure}
It is trivial to see that the connectivity constant $\omega$ for HWs on 3-simplex lattice is equal to 1, whereas
Bradley found that $\omega=1.399710...$ for the 4-simplex \cite{Bradley}. In principle, it is possible to
establish an exact set of recursion relations for the numbers of suitably chosen prerequisite HWs on any
$n$-simplex, as Bradley did for $n=4$. Here we'll demonstrate the method on $n=5$ and $n=6$ cases.

\subsection{5-simplex}\label{section:omega_5-simplex}

Any closed HW of order ($l+1$) on this fractal can be decomposed
into 5 open HWs through 5-simplices of order $l$. There are two
possible types of these open HWs, as depicted in
Figure~\ref{fig:5simplex}(a). The first corresponds to walks
which enter the simplex at one corner, visit all vertices inside it,
and leave it - we shall denote the number of these walks by
$C_{1,l}$. A walk of the second type enters the simplex at one of
its five corners, wanders around it visiting some of the vertices
inside it, leaves it through the second corner, and afterwards
enters it again at the third corner, visits all the remaining
vertices, and finally leaves it - let the total number of these
walks on the $l$-th order 5-simplex be $C_{2,l}$. The total number
$C_{l+1}$ of closed HWs is equal to
\begin{equation}
C_{l+1}=12 C_{1,l}^5+30C_{1,l}^4C_{2,l}+60 C_{1,l}^3C_{2,l}^2+132 C_{2,l}^5\, \label{eq:5simplex}
\end{equation}
{which we found by computer enumeration, together with the recursion
relations}
\begin{eqnarray}
\fl C_{1,l+1}=6 C_{1,l}^5+30 C_{1,l}^4C_{2,l}+78 C_{1,l}^3C_{2,l}^2+96 C_{1,l}^2C_{2,l}^3+132 C_{1,l}C_{2,l}^4+132
C_{2,l}^5\, ,\nonumber \\
 \fl C_{2,l+1}=2 C_{1,l}^5+13 C_{1,l}^4
C_{2,l}+32 C_{1,l}^3C_{2,l}^2+88 C_{1,l}^2C_{2,l}^3+220
C_{1,l}C_{2,l}^4+186 C_{2,l}^5\, . \label{eq:recur5simplex}
\end{eqnarray}
Initial values for these numbers are $C_{1,1}=6$ and $C_{2,1}=2$.
{From the recursion relation (\ref{eq:5simplex}) it follows that
\begin{equation} \frac{\ln C_{l+1}}{5^{l+1}}=\frac{\ln
C_{2,l}}{5^l}+
 \frac{1}{5^{l+1}}\left(132+60 x_l^3+30 x_l^4+12 x_l^5 \right),
\label{eq:kvadrat}
\end{equation}
where the new variable $x_l={{C_{1,l}}/{C_{2,l}}}$ satisfies the
recursion relation
\begin{equation}
x_{l+1}= 6{\frac{ 22 + 22\,x_l + 16\,{x_l^2} + 13\,{x_l^3} +
5\,{x_l^4} + {x_l^5}
       }{186 + 220\,x_l + 88\,{x_l^2} + 32\,{x_l^3} + 13\,{x_l^4} +
       2\,{x_l^5}}}\, ,  \label{eq:recx5simplex}
\end{equation}
obtained from (\ref{eq:recur5simplex}). Numerically iterating this
relation, starting with $x_1=3$, we find that $x_l\to
0.802318837...$ when $l\to\infty$, and consequently, since
$N_l=5^l$, it follows that
\[
\fl \ln\omega=\lim_{l\to\infty}{{\ln
C_{l+1}}\over{5^{l+1}}}=\lim_{l\to\infty}{{\ln
C_{2,l}}\over{5^l}}+\lim_{l\to\infty}{1\over{5^{l+1}}}\ln\left(1+{5\over{11}}x_l^3+
{5\over{22}}x_l^4+{1\over{11}}x_l^5\right)=\lim_{l\to\infty}{{\ln
C_{2,l}}\over{5^l}}\, .
\]
The last limiting value} can be quickly calculated if we introduce
the variable
\[
y_l={{\ln C_{2,l}}\over{5^l}}-{{\ln{186}}\over 4}\left({1\over
5}-{1\over{5^l}}\right)-{{\ln 2}\over 5}\, ,
\]
which, according to (\ref{eq:recur5simplex}), obeys the recursion
relation
\begin{equation}
y_{l+1}=y_l+{1\over{5^{l+1}}}\ln\left(1+{{110}\over{93}}x_l+{{44}\over{93}}x_l^2+
{{16}\over{93}}x_l^3+{{13}\over{186}}x_l^4+{1\over{93}}x_l^5\right)\,
\label{eq:recy5simplex} ,
\end{equation}
and has the initial value $y_l=0$. Then, iterating
(\ref{eq:recy5simplex}) simultaneously with (\ref{eq:recx5simplex})
we find $y_l\to 0.141065489481...$ for large $l$, and finally, since
$\ln\omega=\lim_{l\to\infty}y_l+\ln 186/20+\ln 2/5$, {we obtain
$\omega=1.717769...$.}

To examine the leading order correction to the number of Hamiltonian
walks on the 5-simplex fractal, we note that for any $k$ the
quantity $y_l$ can be written {as
$y_l=y_k+\sum_{m=k}^{l-1}(y_{m+1}-y_m)$. Taking the $l \to \infty$
limit and keeping $k$ fixed in that equation}, one obtains:
\[
\fl \ln \omega=\frac{\ln C_{2,k}}{5^k}+\frac{\ln 186}{4\cdot 5^k
}+\sum _{m=k}^{\infty} \frac{1}{5^{m+1}}\ln \left (
1+\frac{110}{93}x_m+ \frac{44}{93}x_m^2+\frac{16}{93}
x_m^3+\frac{13}{186}x_m^4+\frac{1}{93} x_m^5 \right )\,.\] Since
$x_1=3$ and the array $x_m$ is monotonically decreasing, the sum on
the right hand side of the above equation is bounded from above by
\[
\ln (21.72) \sum_{m=k}^{\infty} \frac{1}{5^{m+1}}=\ln (21.72)
\frac{1}{5^{k+1}} (1+\frac{1}{5}+...) =\frac{1}{4} \ln (21.72)
\frac{1}{5^k}.
\]
Therefore we can conclude that
\begin{equation}
\fl \ln C_{2,k}=5^k \ln \omega+O(1)\,  \qquad \mathrm{and} \qquad
\ln C_{1,k}=\ln C_{2,k}+\ln x_k   =5^k \ln \omega+O(1).
\label{eq:C2_5-simplex_corr}
\end{equation}
Finally, for the {number $C_l$ of closed HWs, from
(\ref{eq:kvadrat}), we obtain the same behavior:}
\begin{equation}
\ln C_{l}=5^l \ln \omega +O(1).  \label{eq:trecakorekcija}
\end{equation}
Comparing with equation (\ref{eq:asimptotika}) we see that in the
case of the 5-simplex fractal both the surface and the power
correction to the number of HWs are absent.

\subsection{6-simplex}{\label{subsec:6simplex}}

In addition to the $C_{1,l}$ and $C_{2,l}$-type walks, already
defined for the 5-simplex, one should introduce one other type of
walks for complete enumeration of HWs on the 6-simplex lattice. These
walks  enter the 6-simplex three times, as depicted in Figure~
\ref{fig:5simplex}(b). Let's denote their number  on the 6-simplex
of order $l$ by $C_{3,l}$. The total number $C_{l+1}$ of closed HWs
within the 6-simplex of order $(l+1)$ is equal to
\begin{eqnarray}
\fl C_{l+1}=60\ C_{1,l}^6+360\ C_{1,l}^5C_{2,l}+1170\
C_{1,l}^4C_{2,l}^2+1920\ C_{1,l}^3C_{2,l}^3  +3960\
C_{1,l}^2C_{2,l}^4\nonumber\\+7920\ C_{1,l}C_{2,l}^5+5580\ C_{2,l}^6
\, , \label{eq:6simplex}
\end{eqnarray}
as we found by computer enumeration. Numbers $C_{1,l}$, $C_{2,l}$
and $C_{3,l}$ satisfy closed set of recursion relations, which can
be numerically analyzed in a way similar, although more complicated
than that used in the case of the 5-simplex lattice (see \ref{Dodatak6simplex}). In
contrast to the 5-simplex case, where numbers $C_{1,l}$, $C_{2,l}$
and $C_l$ have the same asymptotic behaviour
(\ref{eq:C2_5-simplex_corr}) and (\ref{eq:trecakorekcija}), here we
obtained
\begin{eqnarray}
 \ln C_{1,l}&=& 6^l \ln \omega +2^{l+1} \ln \lambda+O(1) \, , \label{eq:47}\\
\ln C_{2,l}&=& 6^l \ln \omega+2^l \ln \lambda+O(1)\, ,\label{eq:48}\\
 \ln C_{3,l}&=&6^l \ln \omega +O(1)\, ,\label{eq:49}
\end{eqnarray}
with $\omega=2.0550...$, $\lambda=0.9864$ and
\begin{equation}
\ln C_l=N_l \ln \omega +N_l^\sigma\ln \mu_S  +O(1)\, ,
\label{eq:410}
\end{equation}
with $N_l=6^l$, $\sigma=\ln 2/\ln 6$ and $\mu_S=\lambda^3$. The
surface correction to the number of HWs is present here, but still
there is no term proportional to $\ln N_l$, which would imply the
presence of the power correction. The same conclusion was obtained
by Bradley \cite{Bradley} for the 4-simplex lattice, with
$\sigma=1/2$.

\section{Discussion and conclusions}

\begin{figure}
\begin{center}
\includegraphics[width=90mm]{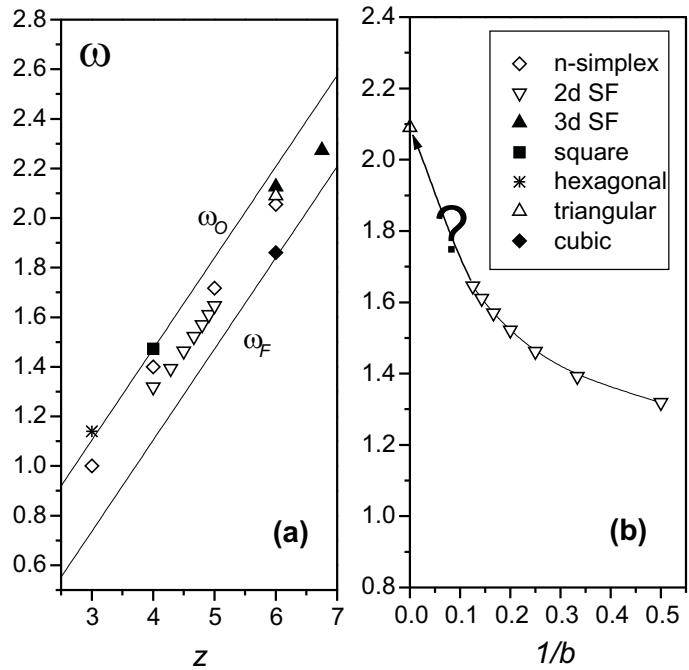}
\end{center}
\caption{(a) Connectivity constant $\omega$ for Hamiltonian walks
on Sierpinski fractals ($\opentriangledown$\/  for 2d SF, and full
triangles for 3d SF) and $n$--simplex lattices ($\diamondsuit$) as
functions of the coordination number $z$ together with previously
found results for 4-simplex [13], hexagonal [12] (star), square
[29] ($\fullsquare$\/), triangular [30] ($\triangle$), and cubic
[15] (full diamond) lattices, as well as Flory [28] and Orland
[21] approximations, $\omega_F=(z-1)/e$ and $\omega_O=z/e$,
respectively. (b) Connectivity constant for 2d SFs as function of
the reciprocal of the scaling parameter $b$, where line connecting
triangles serves merely as the guide to the eye. Open up-oriented
triangle on vertical axes depicts the value of $\omega$ for
triangular lattice.} \label{fig:diskusija}
\end{figure}

The values of the connectivity constant $\omega$ found by us, as
well as some values previously found by other authors \cite{Bradley}
are depicted in Figure~\ref{fig:diskusija}(a), as functions of the
coordination number $z$\footnote{The coordination number $z$ (the
average number of nearest neighbours per site) for the $n$-simplex
lattice is equal to $n$, whereas it can be shown that
$z=6(b+2)/(b+4)$ for 2d SF with the scaling parameter $b$, $z=6$ for
$b=2$ 3d SF  and $z=6.75$ for $b=3$ 3d SF.} of the lattice, together
with the Flory \cite{Flory} and Orland \cite{Orland} approximations,
and $\omega$ for hexagonal \cite{Batchelor1}, square \cite{Jane},
triangular \cite{Batchelor2} and cubic \cite{Pande} lattices.  One
can clearly see that  $\omega$ increases with $z$, which is in
accord with Flory $\omega_F=(z-1)/e$ and Orland {\em et al\/}
$\omega_O=z/e$ formulas, but it is obvious that $\omega$ depends on
other lattice properties too. Furthermore, all values of $\omega$
lie between the values predicted by these two formulas, and it seems
that these kind of approximations give satisfying results for
fractal lattices studied here. Based on the information in this figure
one may conclude that $\omega_O$ and $\omega_F$ are good upper and
lower bounds for fractal HW connectivity constants.

In Figure~\ref{fig:diskusija}(b)
the connectivity constant $\omega$ for 2d SFs is presented as
a function of the reciprocal of the fractal scaling parameter $b$. It
seems that for $b\to\infty$ the connectivity constant might
approach its triangular lattice value. This is not surprising,
since for $b=\infty$ the first step of the construction of the
corresponding 2d SF is already the wedge of the triangular
lattice. In the same limit, 3d SFs approach the corresponding
three-dimensional Euclidean lattice, so it would be useful to
obtain recursion relations for HWs on these fractals for larger
$b$, since there are fewer results for more realistic
three-dimensional lattices. This task requires faster computers,
as well as establishing a better algorithm for enumerating the HW
configurations within a generator of a fractal, which is something
we are planning to do in the nearest future.

As for the correction to the leading-order asymptotic behavior of
the number of HWs, we have shown that the surface correction
$\mu_s^{N^\sigma}$ appears for neither the two-dimensional
Sierpinski fractals nor for the 5-simplex lattice. In contrast, for
both three-dimensional Sierpinski fractals  considered here, as well
as for the 4-simplex \cite{Bradley} and the 6-simplex lattice, the
surface correction is present, with the value of the exponent
$\sigma=1/d_f$, where $d_f$ is the fractal dimension of the
corresponding lattice\footnote{Fractal dimension for 3d SF is
$d_f=\ln(b(b+1)(b+2)/6)/\ln b$, and for the $n$-simplex lattice
$d_f=\ln n/\ln 2$.}. This result is certainly not a simple
generalization of the formula proposed for the homogeneous lattices:
$\sigma=(d-1)/d$. This is actually not surprising: The correction
term $\mu_S^{N^\sigma}$ in (\ref{eq:asimptotika}) was originally
introduced in order to take into account possible surface tension
effects, since at low temperatures a SAW forms a compact globule
(see, for instance, \cite{Owczarek}). The value $\sigma=(d-1)/d$ for
homogeneous lattices then follows from the fact that the surface of
such a globule is proportional to $N^{(d-1)/d}$. In the case of
fractal lattices it is, however, questionable whether such surface
effects exist at all. For instance, all sites of the $b=2$ 3d SF lie
on the surface, which is not the case for the $b=3$ 3d SF, but for
both lattices the number of HWs has the correction term
$\mu_s^{N^\sigma}$. Or, in the case of an $n$-simplex lattice {all
sites have the same number $n$ of neighbors}, and one should not
expect any surface correction, but still, for
some of them the correction $\mu_s^{N^\sigma}$ was found. So, it
seems that {the term $\mu_S^{N^{1/d_f}}$}, obtained for some of the
studied fractals, does not originate from the surface effects, which
is also supported by the fact that $N^{1/d_f}$ is proportional to
the mean radius of the globule formed by a HW, and not to its
surface.

It is interesting to note here that {the existence of the term
$\mu_S^{N^{1/d_f}}$ in the scaling form for the number of HWs},
coincides with the existence of the polymer coil to globule
transition on the corresponding lattice. As was shown earlier, a
polymer chain in a solvent, modeled by SAWs on the 2d Sierpinski
fractals \cite{EKM} and 5-simplex \cite{KumarSingh} can exist only
in the swollen phase; on the 3d Sierpinski fractals
\cite{DharVannimenus,KnezevicVannimenus,EZM}, 4-simplex \cite{Dhar}
and 6-simplex \cite{KumarSingh} lattices, when the temperature is
lowered the polymer undergoes a collapse transition from an expanded
state to a globule state (compact or semi-compact
\cite{KnezevicVannimenus}). Analyzing the asymptotic behaviour of
different types of open HWs on fractals one can observe that the
collapse transition exists on lattices whose topology allows for the
statistical domination of HWs which are not localised. In
particular, HWs on the 2d SG fractals cannot enter a generator of
any order more than once{, i.e. all the walks are localised. On the
5-simplex fractal this is possible (see
Figure~\ref{fig:5simplex}(a)), but the number $C_{1,l}$ of localised
HWs and the number $C_{2,l}$ of delocalised HWs (walks that enter
every $l$th order 5-simplex twice) are of the same order (see
(\ref{eq:C2_5-simplex_corr}), i.e. the delocalised HWs do not
dominate. On the other hand, the delocalised} HWs on the 3d SG
fractals ($j$-type, see Figure~\ref{fig:3dSG}) and on the 6-simplex
($C_{2,l}$-- and $C_{3,l}$--type, see Figure~\ref{fig:5simplex}(b)),
as well as on the 4--simplex \cite{Bradley}, are possible, and
furthermore, the number of these walks is much larger than the
number of the localised HWs (see section~\ref{sec:3dSG} and
subsection~\ref{subsec:6simplex}). This observation strongly
resembles the conclusion obtained in a series of recent papers
\cite{Orlandini} about the delocalisation of knots in the
low-temperature globular phase. Although the term 'delocalisation'
was not used in quite the same sense in these two contexts, it seems
that the same effect is in question, and this problem deserves
further investigation.

Finally, the power dependence of the overall number $C_N$ of closed
HWs on the number of sites $N$ of the lattice was found only for the
3d SFs. A more detailed inspection of the calculation of the
exponent $a$ for these lattices reveals {that
\[
a =\mathrm{const} \lim_{l\to\infty}{{\ln{{h_l}\over{i_l}}}\over l}\,
.
\]
The} numbers $h_l$ and $i_l$ correspond to the localised open
Hamiltonian configurations that visit the maximal (4) and the minimal (2)
number of vertices, respectively, {within} the generator of order
$l$ (see Figure~\ref{fig:3dSG}). In the case of 2d SFs, numbers of
open configurations visiting the maximal or the minimal number of vertices
were $h_l$ and $g_l$, respectively, and it was shown that the ratio
$h_l/g_l$=const  for every $l$. {Consequently, $\lim_{l\to\infty}\ln
(h_l/i_l)/l=0$}, which may be the formal explanation for the absence
of the power correction to the number of HWs on 2d SF. On the other
hand, on the $n$-simplex lattices only one type of localised HW
configurations is possible, so it appears that the power term in the
scaling form for the number of HWs is obtained on lattices where a
larger number of different types of localised configurations is
possible.

In conclusion we can say that the method of exact recursion relations
turned out to be very powerful for the generation and the
enumeration of extremely long Hamiltonian walks on the two- and
three-dimensional Sierpinski and $n$-simplex fractals. Furthermore,
it allows for a detailed numerical analysis of HWs of different
topologies. This enabled us to find various scaling forms for the
number of closed HWs on these lattices. In the case of two-dimensional
Sierpinski fractals, a closed-form expression is obtained for the connectivity
constant. Very  interesting
results were obtained for the three-dimensional Sierpinski fractals.
This should be utilized for attaining deeper insight into the
realistic physical problems which can be modeled by Hamiltonian
walks.

\ack We thank Mr Antun Bala\v z and the Institute of Physics, Belgrade, for providing the computing facilities at
the initial stage of this work. SEH acknowledges the financial support from the Serbian Ministry of Science and
Environmental Protection (Project No: 1634) and JS from the United
States Department of Energy
under contract to the University of California.

\appendix

\section{Recursion relations for HWs on 2d Sierpinski fractals}

In this appendix we present the derivation of relations (\ref{eq:zatvorene2dSF}) and (\ref{eq:otvorene2dSF}) for
open and closed HWs on 2d SFs, for general $b$.

It is obvious that the number of closed HWs on the $(l+1)$th order
generator is of the form $
C_{l+1}=\sum_{i}B_ih_l^{\alpha_i}g_l^{\beta_i}$, where $B_i$ is the
number of all closed Hamiltonian configurations consisting of
${\alpha_i}$ steps of $h$-type and $\beta_i$ steps of $g$-type
("step" is here a part of HW that traverses $l$th order generator
within the considered $(l+1)$th order generator). By definition,
closed HW visits all sites of the $(l+1)$th generator, consequently
it traverses all $b(b+1)/2$ $l$th order generators within it, so
that
\begin{equation}
\alpha_i+\beta_i={{b(b+1)}\over 2}\, . \label{eq:prva}
\end{equation}
On the other hand, every $h$-type step occupies all three vertices of the $l$th generator it traverses, whereas
a $g$-type step occupies only two vertices (the entering and the
exiting one). This means that the numbers ${\alpha_i}$
and $\beta_i$ also have to satisfy the following equation
\begin{equation}
2\alpha_i+\beta_i={{(b+1)(b+2)}\over 2},  \label{eq:druga}
\end{equation}
since the number of  $l$th order vertices inside the ($l+1$)th
generator is $(b+1)(b+2)/2$. The system of equations (\ref{eq:prva})
and (\ref{eq:druga}) has a unique solution $\alpha_i=b+1$,
$\beta_i=(b+1)(b-2)/2$ which completes the derivation of the
relation (\ref{eq:zatvorene2dSF}).

For open $h$-type HWs, in a similar way, we have
$h_{l+1}=\sum_{i}A_ih_l^{x_i}g_l^{y_i}$, where exponents $x_i$ and
$y_i$ satisfy the system $x_i+y_i={{b(b+1)}\over 2}$,
$2x_i+y_i={{(b+1)(b+2)}\over 2}-1$, whose only solution is $ x_i=b$,
$y_i={{b(b-1)}/2}$. At the same time, every $g$-type HW on a
generator $G_{l+1}^2$ can be obtained from one and only one $h$-type
HW (and vice versa), by substituting one $h$-step with a $g$-step,
as depicted in Figure~A1.
\begin{figure}
\begin{center}
\includegraphics[width=50mm]{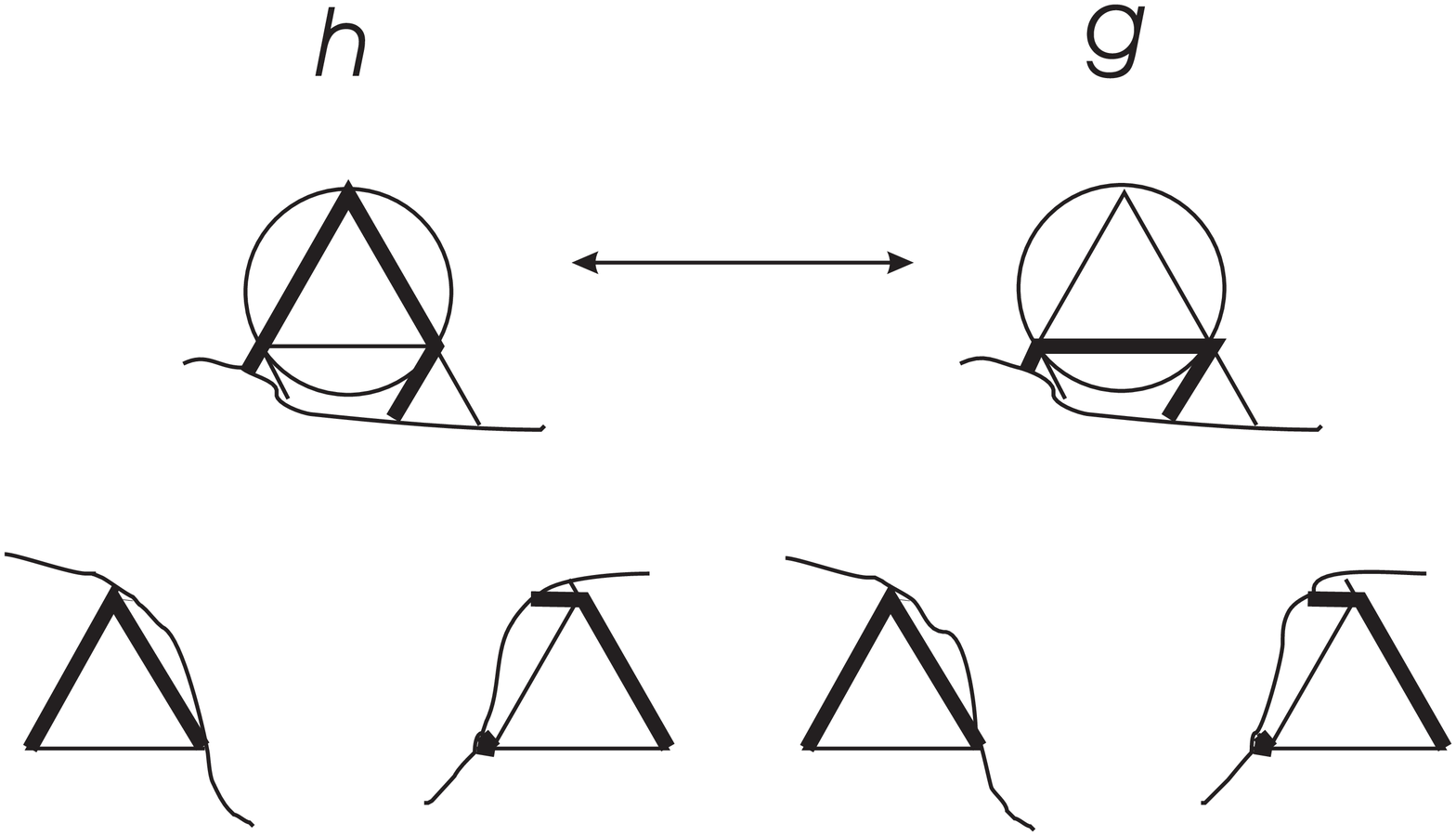} \label{fig:dodatak}
\end{center}
\caption{$h-$ and $g-$type HW on a $G_{l+1}^2$ generator, which
can be obtained one from another in a unique way. These two walks
differ only in the parts surrounded by circles.}
\end{figure}
This means that all of these walks have exactly $(b-1)$ $h$-steps and $[b(b-1)/2+1]$ $g$-steps, i.e. relations
(\ref{eq:otvorene2dSF}) are correct.

\section{RG equations for SAW model on 2d SF}

In this appendix we give exact RG equations (\ref{eq:recurB}) for
calculating the connectivity constant for the ordinary SAW model on
the $b=4$ and the $b=5$ two-dimensional Sierpinski fractals. These equations were
obtained via computer enumeration and classification of all SAW
configurations within the corresponding fractal generator. RG
relations (\ref{eq:recurB}) for $b=6$ and 7 were also found, but
they are too cumbersome to be quoted here and they are available
upon request.

\begin{eqnarray}
\lo b=4:  \nonumber \\ \fl B'=
  {B^4} + 6{B^5} + 6{B^6} + 9{B^7} + 9{B^8} + 9{B^9} +
   4{B^{10}} + 4{B^3}B_1 + 30{B^4}B_1 +36{B^5}B_1 \nonumber \\
\fl \qquad  + 56{B^6}B_1 +
   58{B^7}B_1 + 56{B^8}B_1 + 26{B^9}B_1 +
 6{B^2}{B_1^2} + 57{B^3}{B_1^2} +84{B^4}{B_1^2}  \nonumber\\
\fl \qquad + 134{B^5}{B_1^2} +
   143{B^6}{B_1^2} +
    128{B^7}{B_1^2} +56{B^8}{B_1^2} + 4B{B_1^3} +51{B^2}{B_1^3} + 96{B^3}{B_1^3}
   \nonumber \\
\fl \qquad  + 156{B^4}{B_1^3} +168{B^5}{B_1^3} + 128{B^6}{B_1^3} + \underline{40{B^7}{B_1^3}} + {B_1^4} +
   21B{B_1^4} + 55{B^2}{B_1^4}
    \nonumber \\
\fl \qquad +93{B^3}{B_1^4} +94{B^4}{B_1^4} +48{B^5}{B_1^4} + 3{B_1^5} +14B{B_1^5} + 28{B^2}{B_1^5}\nonumber \\
\lo{+}
   20{B^3}{B_1^5} + {B_1^6} + 4B{B_1^6}
      \nonumber \\
\fl B_1' = {B^6}B_1 + 6{B^7}B_1 + 7{B^8}B_1 + 4{B^9}B_1 + 6{B^5}{B_1^2} +
   38{B^6}{B_1^2} +44{B^7}{B_1^2} +  26{B^8}{B_1^2}\nonumber\\
   \fl\qquad +
   14{B^4}{B_1^3} + 92{B^5}{B_1^3} + 102{B^6}{B_1^3}  +
   56{B^7}{B_1^3} +   16{B^3}{B_1^4} +106{B^4}{B_1^4} +
   104{B^5}{B_1^4}  \nonumber\\
\lo{+} \underline{40{B^6}{B_1^4}} + 9{B^2}{B_1^5} +
    58{B^3}{B_1^5} +40{B^4}{B_1^5} + 2B{B_1^6} + 12{B^2}{B_1^6}
\nonumber\\
\lo b=5:
 \nonumber \\
\fl B'=
 {B^5} + 10{B^6} + 20{B^7} + 30{B^8} + 54{B^9} + 68{B^{10}} +
   98{B^{11}} + 94{B^{12}} +86{B^{13}} \nonumber\\
\fl \qquad + 38{B^{14}} + 16{B^{15}} +   5{B^4}B_1 + 60{B^5}B_1 + 140{B^6}B_1 +228{B^7}B_1 +
  443{B^8}B_1 \nonumber\\
   \fl \qquad + 586{B^9}B_1 + 867{B^{10}}B_1 +
854{B^{11}}B_1 +786{B^{12}}B_1+ 348{B^{13}}B_1 + 140{B^{14}}B_1 \nonumber \\
 \fl \qquad
     +
   10{B^3}{B_1^2} + 146{B^4}{B_1^2} + 402{B^5}{B_1^2} +
   718{B^6}{B_1^2}+ 1521{B^7}{B_1^2} + 2137{B^8}{B_1^2}  \nonumber\\
\fl \qquad +
   3203{B^9}{B_1^2} +3240{B^{10}}{B_1^2} + 2918{B^{11}}{B_1^2} +
   1268{B^{12}}{B_1^2} + 458{B^{13}}{B_1^2} + 10{B^2}{B_1^3}\nonumber\\
\fl \qquad +
   184{B^3}{B_1^3} + 610{B^4}{B_1^3} + 1218{B^5}{B_1^3} +
   2846{B^6}{B_1^3} +4316{B^7}{B_1^3} +6433{B^8}{B_1^3} \nonumber\\
\fl \qquad +
   6648{B^9}{B_1^3} + 5630{B^{10}}{B_1^3} + 2306{B^{11}}{B_1^3} +
   664{B^{12}}{B_1^3} + 5B{B_1^4} + 126{B^2}{B_1^4} \nonumber \\
\fl \qquad +
   523{B^3}{B_1^4} + 1209{B^4}{B_1^4}  +3170{B^5}{B_1^4} +
   5307{B^6}{B_1^4} + 7678{B^7}{B_1^4} + 7960{B^8}{B_1^4} \nonumber\\
   \fl \qquad+
   5960{B^9}{B_1^4} + 2104{B^{10}}{B_1^4} + \underline{360{B^{11}}{B_1^4}} +
   {B_1^5} + 44B{B_1^5} + 249{B^2}{B_1^5} + 710{B^3}{B_1^5}
    \nonumber \\
\fl \qquad +2159{B^4}{B_1^5} + 4118{B^5}{B_1^5} + 5604{B^6}{B_1^5} +
   5554{B^7}{B_1^5} + 3292{B^8}{B_1^5} +776{B^9}{B_1^5}\nonumber\\
   \fl \qquad + 6{B_1^6} +
   59B{B_1^6} + 234{B^2}{B_1^6} + 891{B^3}{B_1^6} + 2031{B^4}{B_1^6}+ 2479{B^5}{B_1^6} + 2086{B^6}{B_1^6}  \nonumber \\
\fl \qquad +
   744{B^7}{B_1^6} + 5{B_1^7} + 36B{B_1^7} + 214{B^2}{B_1^7} +
   630{B^3}{B_1^7} + 626{B^4}{B_1^7} + 324{B^5}{B_1^7}\nonumber\\
\lo{+} {B_1^8} +
   28B{B_1^8} +117{B^2}{B_1^8} + 72{B^3}{B_1^8} + 2{B_1^9} +
   10B{B_1^9}\nonumber\\
\fl B_1' = {B^8}B_1 + 12{B^9}B_1 + 39{B^{10}}B_1 + 48{B^{11}}B_1 +
   60{B^{12}}B_1 + 34{B^{13}}B_1 +  16{B^{14}}B_1\nonumber \\
\fl \qquad + 8{B^7}{B_1^2}  +
   102{B^8}{B_1^2} + 344{B^9}{B_1^2} + 432{B^{10}}{B_1^2} +  556{B^{11}}{B_1^2} + 314{B^{12}}{B_1^2}  \nonumber \\
\fl \qquad   + 140{B^{13}}{B_1^2}
 +
   27{B^6}{B_1^3} + 366{B^7}{B_1^3} +  1278{B^8}{B_1^3} +
   1616{B^9}{B_1^3} + 2098{B^{10}}{B_1^3}  \nonumber \\
\fl \qquad + 1156{B^{11}}{B_1^3} +
   458{B^{12}}{B_1^3} + 50{B^5}{B_1^4} + 722{B^6}{B_1^4} +
   2600{B^7}{B_1^4} + 3254{B^8}{B_1^4}  \nonumber \\
\fl \qquad + 4128{B^9}{B_1^4} +  2128{B^{10}}{B_1^4} + 664{B^{11}}{B_1^4}
   + 55{B^4}{B_1^5} +
   852{B^5}{B_1^5} + 3148{B^6}{B_1^5} \nonumber \\
   \fl \qquad + 3808{B^7}{B_1^5} +
   4474{B^8}{B_1^5} + 1968{B^9}{B_1^5} + \underline{360{B^{10}}{B_1^5}}+
   36{B^3}{B_1^6}+ 610{B^4}{B_1^6}  \nonumber \\
\fl \qquad  + 2302{B^5}{B_1^6} +
   2590{B^6}{B_1^6} + 2540{B^7}{B_1^6} +736{B^8}{B_1^6} +
   13{B^2}{B_1^7}+ 254{B^3}{B_1^7}  \nonumber \\
\fl \qquad  + 981{B^4}{B_1^7} +
   948{B^5}{B_1^7} +  592{B^6}{B_1^7}  +2B{B_1^8} + 54{B^2}{B_1^8} +
   220{B^3}{B_1^8}  \nonumber\\
\lo{+}144{B^4}{B_1^8} + 4B{B_1^9} + 20{B^2}{B_1^9} \, .
\end{eqnarray}

\section{Recursion relations for HWs within the $b=3$ 3d SF \label{Dodatak3dSF}}

Here we give the recursion relations for the numbers $g_l$, $h_l$,
$i_l$ and $j_l$ of open HW configurations within the $b=3$ 3d SF,
obtained via computer enumeration. With  symbols $g_{l+1}$,
$h_{l+1}$, $i_{l+1}$, $j_{l+1}$, $g_l$, $h_l$, $j_l$ and $i_l$
abbreviated to $g'$, $h'$, $i'$, $j'$, $g$, $h$, $i$ and  $j$,
respectively, these relations have the following form:
\begin{eqnarray}
\fl
g'=6120i^2g^3j^5+3312i^2g^4j^4+8176i^2g^5j^3+5068i^2g^6j^2+2964i^2g^7j+776i^2g^8\nonumber
\\
\fl{\qquad
+}13296i^3gj^5h+12688i^3g^2j^4h+36832i^3g^3j^3h+36504i^3g^4j^2h+27768i^3g^5jh\nonumber
\\
\fl{\qquad
+}9200i^3g^6h+2080i^4j^4h^2+20224i^4gj^3h^2+40464i^4g^2j^2h^2+46064i^4g^3jh^2\nonumber
\\
\lo{+}21856i^4g^4h^2+2848i^5j^2h^3+12192i^5gjh^3+11328i^5g^2h^3+512i^6h^4
\nonumber\\ \fl
h'=2928ig^4j^5+1288ig^5j^4+3296ig^6j^3+1760ig^7j^2+936ig^8j+232ig^9+13296i^2g^2j^5h\nonumber
\\
\fl{\qquad
+}10608i^2g^3j^4h+28160i^2g^4j^3h+23840i^2g^5j^2h+16632i^2g^6jh+5168i^2g^7h\nonumber\\
\fl{ \qquad+}
12768i^3j^5h^2+14336i^3gj^4h^2{+}50176i^3g^2j^3h^2+63232i^3g^3j^2h^2+56624i^3g^4jh^2\nonumber\\
\fl{\qquad
+}21856i^3g^5h^2+13824i^4j^3h^3+34432i^4gj^2h^3+48928i^4g^2jh^3+27968i^4g^3h^3\nonumber\\
\lo{+}5824i^5jh^4+8192i^5gh^4\nonumber \\ \fl
i'=12504i^3g^2j^5+7880i^3g^3j^4+19240i^3g^4j^3+13232i^3g^5j^2+8224i^3g^6j+2184i^3g^7\nonumber\\
\fl{\qquad +}
528i^4j^5h+4592i^4gj^4h{+}24224i^4g^2j^3h+36928i^4g^3j^2h+33728i^4g^4jh{+}13296i^4g^5h\nonumber\\
\fl{\qquad +}
1184i^5j^3h^2+8384i^5gj^2h^2+18880i^5g^2jh^2{+}13024i^5g^3h^2+640i^6jh^3+1600i^6gh^3\nonumber
\\ \fl
j'=4308i^2j^8+5808i^2gj^7+17424i^2g^2j^6+11936i^2g^3j^5+19164i^2g^4j^4{+}14096i^2g^5j^3\nonumber\\
\fl{ \qquad+}
9208i^2g^6j^2+3360i^2g^7j+544i^2g^8+11616i^3j^6h{+}21440i^3gj^5h+51024i^3g^2j^4h\nonumber\\
\fl{ \qquad +}
66096i^3g^3j^3h+56056i^3g^4j^2h{+}28864i^3g^5jh+6400i^3g^6h+10312i^4j^4h^2\nonumber\\
\fl{ \qquad +} 35296i^4gj^3h^2
{+}53248i^4g^2j^2h^2+45440i^4g^3jh^2+14080i^4g^4h^2+5728i^5j^2h^3\nonumber\\
\lo{+}12544i^5gjh^3+7168i^5g^2h^3+512i^6h^4 \, .\nonumber
\end{eqnarray}
The initial values of these numbers are $g_1=497000$, $h_1=728480$,
$i_1= 340476$,  $j_1=811468$ and the formula for the number $
C_{l+1}$ of closed HWs within $G_{l+1}^3(3)$ is:
\begin{eqnarray}
\fl C_{l+1}=92g_l^8j_l^2+48g_l^9j_l+8g_l^{10}+1792i_l g_l^6 h_l
j_l^2+1248 i_l g_l^7 h_l j_l+384 i_l g_l^8 h_l\nonumber\\ \fl{
\qquad +}7568 i_l^2 g_l^4 h_l^2 j_l^2  +7104i_l^2 g_l^5 h_l^2
j_l+3008 i_l^2g_l^6h_l^2+10560 i_l^3 g_l^2 h_l^3 j_l^2+13440
i_l^3g_l^3h_l^3j_l\nonumber\\ \fl{ \qquad +}7680i_l^3g_l^4h_l^
3+4480i_l^4h_l^4j_l^2+7680i_l^4 g_l h_l^4 j_l+6016 i_l^4 g_l^2 h_l^4
+ 512 i_l^5 h_l^5 \, .\nonumber
\end{eqnarray}

\section{Analysis of recursion relations for HWs on 6-simplex lattice \label{Dodatak6simplex}}

Recursion relations for the numbers $C_{1,l}$, $C_2$, and $C_{3,l}$
of open HW configurations within the 6-simplex of order $l$ have the
following form:
\begin{eqnarray}
\fl C_1'=5544 C_1^2C_2^4+1728 C_1^3 C_2^2 C_3+ 2592 C_1^3
C_2^3+120 C_1^4 C_3^2+480 C_1^4C_2C_3+960 C_1^4 C_2^2  \nonumber \\
\fl {\quad +} 48 C_1^5 C_3+216 C_1^5 C_2+24 C_1^6+25008 C_2^4
C_3^2+20544 C_2^5 C_3+ 6576 C_2^6  +11328 C_1C_2^3C_3^2\nonumber
\\ \fl {\qquad +} 15264 C_1C_2^4C_3+8688 C_1 C_2^5{+}
4992 C_1^2 C_2^3 C_3 \label{eq:6simplex1}\\
\fl C_2'=94336 C_2^2C_3^4+76800 C_2^3C_3^3+48160 C_2^4C_3^2+ 23520
C_2^5 C_3+6576 C_1 C_2^5 +17120 C_1C_2^4C_3\nonumber
\\ \fl\quad {+}16672 C_1 C_2^3
C_3^2+2832 C_1^2 C_2^2 C_3^2+ 5088
C_1^2 C_2^3 C_3+832 C_1^3C_2^2 C_3+ 3620 C_1^2 C_2^4 \nonumber\\
\fl \qquad  +1232 C_1^3C_2^3{+} 144 C_1^4 C_2 C_3+324 C_1^4 C_2^2{+}
64 C_1^5 C_2+6 C_1^6+16 C_1^5 C_3\label{eq:6simplex2}
\\
\fl C_3'=541568 C_3^6+94336 C_2^3 C_3^3+43200 C_2^4 C_3^2+
14448 C_2^5 C_3+2940C_2^6+6252 C_1 C_2^4 C_3\nonumber\\
\fl{\quad +}2568 C_1 C_2^5+1416 C_1^2 C_2^3 C_3+954 C_1^2 C_2^4+208
C_1^3 C_2^3{+}54 C_1^4 C_2^2\nonumber\\
\fl\qquad{+}6 C_1^5 C_3+12 C_1^5 C_2+C_1^6\, ,\label{eq:6simplex3}
\end{eqnarray}
where $C_i'=C_{i,l+1}$ and $C_i=C_{i,l}$, with the initial values
$C_{1,1}=24$, $C_{2,1}=6$, $C_{3,1}=1$. Introducing new variables
\[
x_l={{C_{1,l}}\over{C_{2,l}}}\, , \quad
y_l={{C_{2,l}}\over{C_{3,l}}}\, , \quad z_l={{\ln
C_{3,l}}\over{6^l}}-{{\ln 541568}\over 5}\left({1\over
6}-{1\over{6^l}}\right)\, ,
\]
{one can obtain closed set of recursion relations} which iterates
towards $x_l, y_l\to 0, z_l\to 0.280204...$ for $l\to\infty$. On the
other hand, from (\ref{eq:6simplex}) it follows that
\[ \fl
{{\ln C_{l+1}}\over{6^{l+1}}}={{\ln y_l}\over{6^l}}+{{\ln
5580}\over{6^{l+1}}}+{1\over{6^{l+1}}}\ln\left(1+{{44}\over{31}}x_l+{{22}\over{31}}x_l^2+{{32}\over{93}}x_l^3+{{13}\over{62}}x_l^4+{2\over{31}}x_l^5+{1\over{93}}x_l^6\right)\,
,
\]
whereas from (\ref{eq:6simplex2}) one gets
\begin{eqnarray}
\fl {{\ln C_{2,l+1}}\over{6^{l+1}}}={{\ln
C_{3,l}}\over{6^{l}}}+2{{\ln y_l}\over{6^{l+1}}}+{{\ln
94336}\over{6^{l+1}}}+{1\over{6^{l+1}}}\ln\left(1+{{600}\over{737}}y_l+{{1505}\over{2948}}y_l^2+
\right.{{521}\over{2948}}x_ly_l^2+\nonumber\\ \lo{}
{{177}\over{5896}}x_l^2y_l^2+{{735}\over{2948}}y_l^3+{{535}\over{2948}}x_ly_l^3+
{{159}\over{2948}}x_l^2y_l^3+\left.{1\over{1474}}x_l^5y_l^4+{3\over{47168}}x_l^6y_l^4\right)\,
, \nonumber
\end{eqnarray}
and, consequently,
\[
\ln\omega =\lim_{l\to\infty}{{\ln C_{3,l}}\over{6^l}}+{4\over
3}\lim_{l\to\infty}{{\ln y_l}\over{6^l}}\, .
\]
Since the numerical analysis shows that $\ln y_l/6^l\to 0$, we
finally obtain $\omega=2.0550...$

To find the leading order asymptotic behavior of the number of
Hamiltonian walks on the 6-simplex fractal ($C_{1,l}$,$C_{2,l}$,
$C_{3,l}$ and $C_l$), we conduct an analysis similar to the one used
in the case of the 5-simplex. However, we can tell right away that
$C_{1,l}$,$C_{2,l}$ and $C_{3,l}$ will have mutually different
asymptotics, since their ratios $x_l=\frac{C_{1,l}}{C_{2,l}}$ and
$y_l=\frac{C_{2,l}}{C_{3,l}}$ were found to be approaching zero for
large $l$. By iterating the recursion relations for $x_l$ and $y_l$,
one finds that the ratio of $x_l$ and $y_l$ quickly approaches a
constant equal to 1.521868... Keeping only the terms with the lowest
sum of powers in $x_l$ and $y_l$ in the recursion relations, we find
$x_{l+1}=12 \frac{1042 y_l^2}{47168}$, $y_{l+1}=2 \frac{47168
y_l^2}{541568}$, for $l\gg 1$ and the ratio $x_{l+1}/y_{l+1}$ is
indeed 1.521868... From the above equation for $y_{l+1}$, one can
see that
\begin{equation}
\ln y_l\approx 2^l \ln \lambda, \,\,\,\,
\lambda=\mathrm{const}=\lim_{l\to\infty}\frac{\ln
y_l}{2^l}=0.9864... \label{eq:y_l_approx}
\end{equation}

The asymptotic relation  (\ref{eq:49}) can be obtained starting with
$z_l=z_k+\sum_{m=k}^{l-1} (z_{m+1}-z_m)$ and following a procedure
completely analogous to the one used in the 5-simplex case.
Relations (\ref{eq:47}) and (\ref{eq:48}) then follow from
(\ref{eq:49}), (\ref{eq:y_l_approx}) and the fact that $x_l$ and
$y_l$ are proportional in the large $l$ limit. From
(\ref{eq:47})-(\ref{eq:49}) it is apparent that in the limiting case
$C_{1,l}\ll C_{2,l} \ll C_{3,l}$, since $\lambda < 1$. Therefore, it
holds that $C_{l+1}\approx 5580 C_{2,l}^6$, so finally $\ln C_l=6^l
\ln \omega +3*2^l \ln \lambda$, which is equivalent to
(\ref{eq:410}).

\end{document}